\documentclass[12pt, draftclsnofoot, onecolumn]{IEEEtran}
\usepackage{color}
\usepackage{cite}
\usepackage{amsmath}
\usepackage{algorithmic}
\usepackage{algorithm}
\usepackage{array,amsfonts}
\usepackage{setspace}
\usepackage{eqparbox}
\usepackage{amssymb,pst-poly}
\usepackage{bm}
\usepackage{graphicx}
\usepackage{subfigure}
\usepackage{stfloats}
\usepackage{caption}

\usepackage{mathrsfs}
\makeatletter \@addtoreset{equation}{section} \makeatother

\begin{document}

\title{Machine Learning for Intelligent Authentication in 5G-and-Beyond Wireless Networks}

\author{He~Fang,~\IEEEmembership{Student Member, IEEE,}
      Xianbin~Wang,~\IEEEmembership{Fellow, IEEE,}~and~Stefano~Tomasin,~\IEEEmembership{Senior Member, IEEE}

\thanks{H. Fang and X. Wang are with the Department of Electrical and Computer Engineering, The University of Western Ontario, London, ON N6A 5B9, Canada. Email: hfang42@uwo.ca, xianbin.wang@uwo.ca.}

\thanks{S. Tomasin is with the Department of Information Engineering, University of Padova, Padua, Italy, and also with the Consorzio Nazionale Interuniversitario per le Telecomunicazioni, Padova Research Unit, Padua, Italy. Email: tomasin@dei.unipd.it.}
}

\maketitle

\begin{abstract}
The fifth generation (5G) and beyond wireless networks are critical to support diverse vertical applications by connecting heterogeneous  devices and machines, which directly increase vulnerability for various spoofing attacks. Conventional cryptographic and physical layer authentication techniques are facing some challenges in complex dynamic wireless environments,  including significant security overhead, low reliability, as well as difficulties in pre-designing a precise authentication model, providing continuous protection, and learning time-varying attributes. In this article, we envision new authentication approaches based on machine learning techniques by opportunistically leveraging physical layer attributes, and introduce intelligence to authentication for more efficient security provisioning. Machine learning paradigms for intelligent authentication design are presented, namely for parametric/non-parametric and supervised/unsupervised/reinforcement learning algorithms.  In a nutshell, the machine learning-based intelligent authentication approaches utilize specific features in the multi-dimensional domain for achieving cost-effective, more reliable, model-free, continuous, and situation-aware device validation under unknown network conditions and unpredictable dynamics.
\end{abstract}

\IEEEpeerreviewmaketitle

\section*{INTRODUCTION}
The 5G-and-beyond wireless networks have received tremendous attentions in both academia and industries alike, which will enable a wide variety of vertical  applications by connecting heterogeneous devices and machines  \cite{31}. The dramatically growing number of low-cost devices and access points with increased mobility and heterogeneity  leads to the extremely \emph{complex  and dynamic environment} of 5G-and-beyond wireless networks. Specifically, due to the open broadcast nature of radio signal propagation,  widely adopted standardized transmission protocols, and intermittent communication characteristic, wireless communications are extremely vulnerable to interception and spoofing attacks~\cite{01}. As shown in the conventional `Alice-Bob-Eve'  model of Fig. 1,  Alice and Bob are legitimate devices and communicate with each other in the presence of a Spoofer (Eve). This adversary dynamically intends to intercept legitimate communications and to impersonate Alice for obtaining illegal advantages from Bob. The main objective of Bob is to uniquely and unambiguously identify Alice by authentication techniques.

\begin{figure}[htbp]
\centering
\includegraphics[width=12cm,height=6cm]{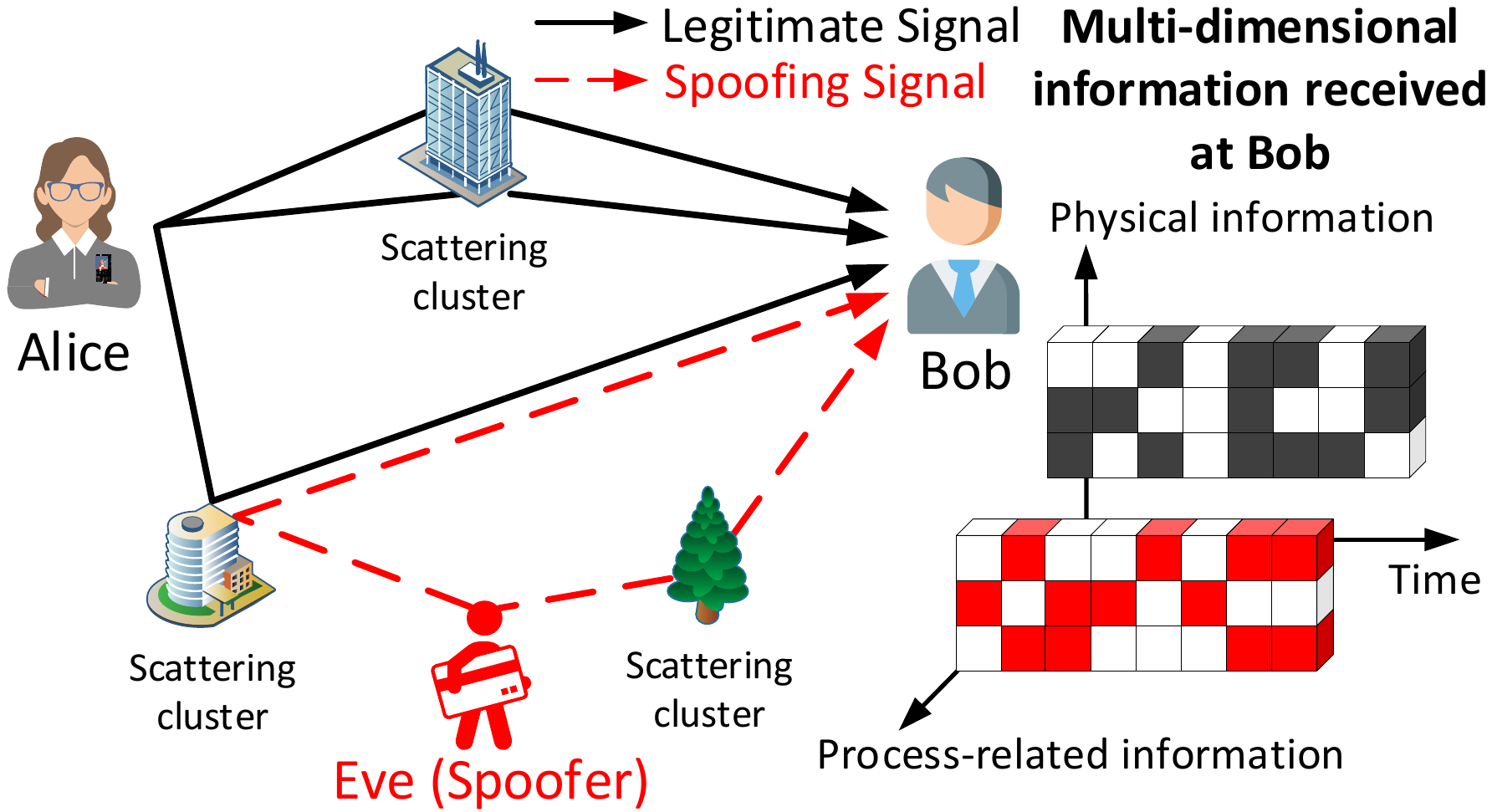}
\caption{Alice-Bob-Eve  model in the complex time-varying environment. Bob identifies Alice from Spoofer based on the multi-dimensional received information, which may be related to communication links, physical environment, and communication process, just to name a few.}
\end{figure}

Although cryptographic techniques have been widely studied for authentication, they may fall short of the desired performance in many emerging scenarios of 5G-and-beyond wireless networks \cite{01}. The fundamental weaknesses of conventional cryptography techniques are the \textbf{increasing latencies, communication and computation overheads} for achieving better security performance, which are extremely undesirable for the delay-sensitive communications and resource-constraint devices. To be specific, appropriate key management procedures are necessary for the conventional cryptographic techniques, and  cooperation among multiple entities is required for both ring and group signature, leading to excessive latencies and significant communication overhead. Furthermore, due to the disregard of inherent features of communicating devices, detecting compromised security keys cannot be readily achieved by the conventional digital credentials-based techniques~\cite{01}.

Physical layer authentication provides an alternative approach of validating a device by exploiting the communication link, device, and location-related attributes, as exemplified by channel impulse response (CIR) \cite{01}, channel frequency response (CFR) in orthogonal frequency division multiplexing (OFDM) \cite{06}, received signal
 strength indicator (RSSI) \cite{01}, carrier frequency offset (CFO) \cite{02}, and radio frequency fingerprint (RFF) \cite{72}.
Such time-varying attributes are inherently related to the unique imperfection of communicating devices and  corresponding environment, which provide specific distinguishing features and natural dynamic protections for legitimate communications.

\section*{CHALLENGES FOR EXISTING PHYSICAL LAYER AUTHENTICATION}
Physical layer authentication has many advantages, such as low computational requirement and network overhead \cite{70,60,82}, but it also faces many challenges. The main reason is that most of the existing physical layer authentication techniques rely on static mechanisms while encountering  the complex and dynamic wireless environment of 5G-and-beyond wireless networks. To be more specific, we summarize the challenges for existing physical layer authentication techniques as follows. \\
\textbf{Low reliability when using single attribute:} Performance of the single attribute-based physical layer authentication schemes suffers from the imperfect estimates and variations of the selected attribute  \cite{01}. Moreover, the limited range of specific attribute distribution may not be sufficient for differentiating transmitters all the time. This limits the performance of single attribute-based authentication schemes in many universal applications, such as mobile device security.\\
\textbf{Difficulty in pre-designing a precise authentication model:} Most of the existing physical layer authentication schemes are model-based, as exemplified by~\cite{02,06}, which may be deteriorated when it is operated in a complex time-varying environment. More importantly, tremendous amounts of data and \emph{a priori} knowledge are required for obtaining the accurate channel model, which are obviously undesirable for those mobile networks, such as unmanned aerial vehicle (UAV) networks and vehicular ad-hoc networks (VANETs). Hence, it is challenging to pre-design a precise authentication model for supporting new applications of 5G-and-beyond wireless networks.  \\
\textbf{Impediment to continuous protection of legitimate devices:} Most of the existing authentication schemes are static in time and binary in nature, which means that the devices either pass or fail the authentication, thus leading to the one-time hard verification~\cite{06,02}.
Due to the intermittent communications between Alice and Bob, these schemes may be limited in detecting Spoofer after the initial login and in addressing varying security risks.  \\
\textbf{Challenge of learning time-varying attributes:} Authentication performance of the existing static authentication schemes could be severely affected by the unpredictable variations of attributes due to the potential decorrelation  at different time instants and device mobility. Hence,  the variations of attributes increase the uncertainty for adversaries, but  decrease the authentication accuracy of legitimate devices operating without learning the diverse attributes as well.

Due to these existing security weaknesses, the authentication enhancement and the efficient security provisioning are of paramount importance for  5G-and-beyond wireless networks, especially in the light of the proliferation of low-cost devices for vertical industrial applications.

\section*{AUTHENTICATION WITH INTELLIGENCE }

In this article, we envision intelligent authentication approaches with the help of machine learning to address the above challenges for security enhancement and more efficient management in 5G-and-beyond networks, which are different from the works of \cite{87,88,89}. As illustrated in Fig. 2, the 5G-and-beyond networks are expected to provide wide services having high communication and security performance as well as privacy preserving transmissions. Based on these basic requirements, the intelligent authentication is required  to meet multi-objectives, namely for high cost efficiency, high reliability, model independence, continuous protection, and situation awareness. More importantly, the flexible designs of security management are extremely helpful in providing automatic authentication in different 5G-and-beyond communication scenarios, which are presented in different colors inside the pentagon of  Fig. 2. Meeting these multi-objectives supports the advantages of intelligent authentication based on machine learning as follows:
\begin{figure}[htbp]
\centering
\includegraphics[width=8cm,height=8cm]{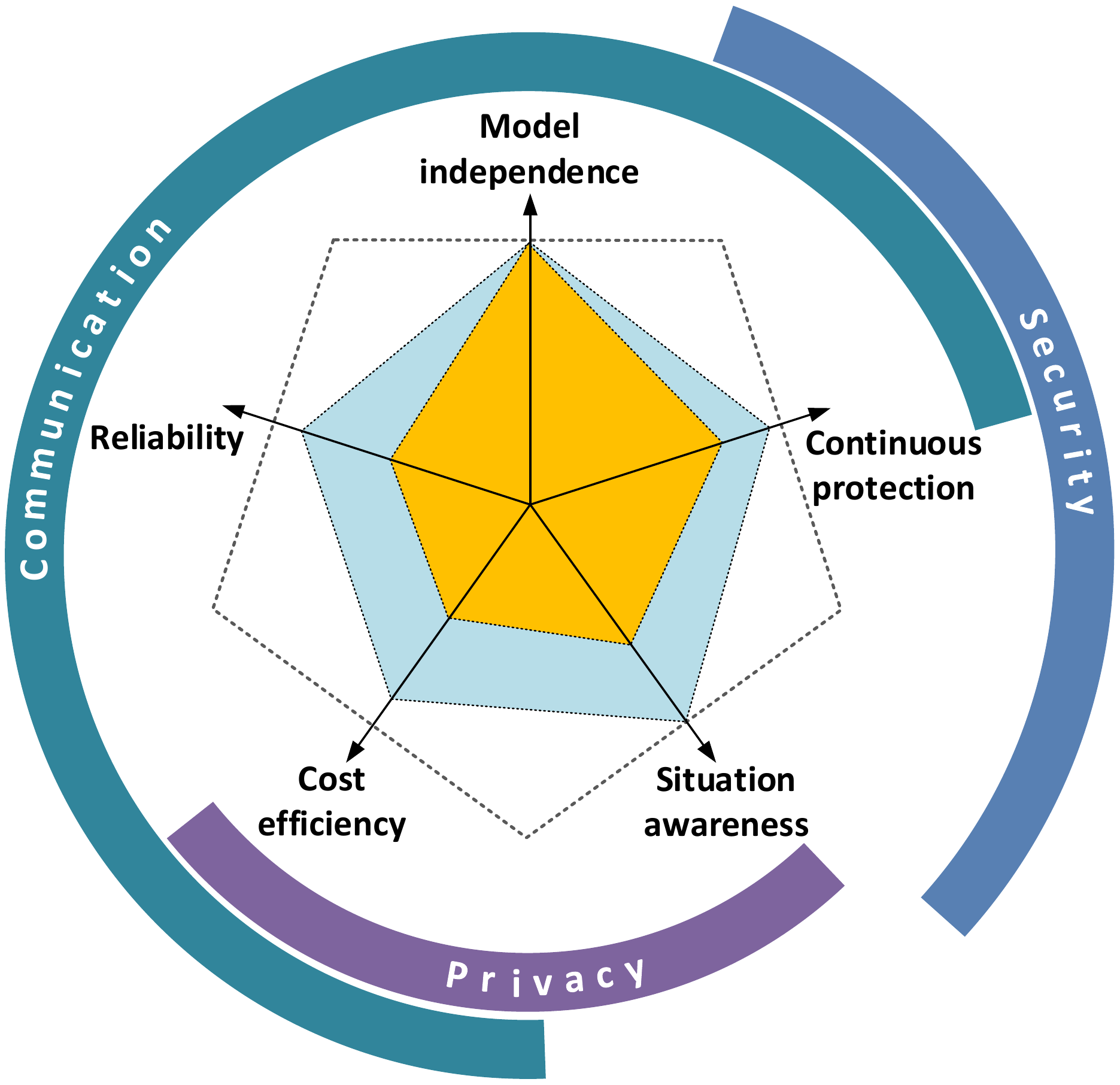}
\caption{Requirements and multi-objectives of intelligent authentication in 5G-and-beyond.}
\end{figure}
\begin{itemize}
\item \textbf{High cost efficiency:}  Due to the increasing number of resource-constraint devices in 5G-and-beyond wireless applications, both communication and security management should be executed concurrently to achieve cost-effective authentication. The opportunistic selection of attributes and dimension reduction methods \cite{01} may help in decreasing the complexity of the authentication system relying on multi-dimensional attributes as well as in reducing  communication and computation overheads. Furthermore, by performing training and testing with devices having enough energy and storage space, simplified and fast authentication can be achieved at resource-constraint devices.
\item \textbf{High reliability:} Utilizing specific features and relationships in the multi-dimensional domain is extremely  helpful for achieving security enhancement, since it is more difficult for an adversary to succeed in predicting or imitating all attributes based on the received signals and observations. To be more specific, the multi-dimensional information, such as time, frequency, network architecture, and communication process, provides broader protections for legitimate users. Machine learning techniques could facilitate the  authentication by analyzing and fusing the multi-dimensional information.
\item \textbf{Model independence:} A data-based scheme through exploring machine learning techniques overcomes the difficulties in modeling uncertainties and unknown dynamics of the authentication process. Hence, the model-free authentication scheme removes the assumption of knowing structures of authentication systems, resulting in a more scalable authentication process design.
\item \textbf{Continuous protection:} Utilizing the received information along with data transmission for authentication could provide identification after login and control the varying security risks. In achieving this, machine learning techniques may be explored for data analysis and processing, so that the seamless protection
for legitimate devices can be achieved in 5G-and-beyond networks.
  Furthermore,  continuous authentication may contribute to the quick authentication handover.
\item \textbf{Situation awareness:} The situation-aware authentication observes the varying security risks and learns from the complex dynamic environment to enhance security. Machine learning techniques provide powerful tools to  learn the dynamic adversarial environment for self-optimization and self-organization, thereafter for achieving the automatic authentication. More importantly, machine learning techniques could help in the detection of time-varying attributes and the adaptation of authentication process.
\end{itemize}

As a conclusion, by exploring machine learning techniques, we introduce the intelligence to authentication in the complex time-varying environment under unknown network conditions and
unpredictable dynamics, supporting radically new applications of 5G-and-beyond wireless networks.

\section*{DESIGN OF MACHINE LEARNING-AIDED INTELLIGENT AUTHENTICATION}

As shown in Fig. 3, we present the design of machine learning-aided intelligent authentication approaches  by utilizing multi-dimensional attributes and by optimizing the holistic authentication process.
\begin{figure}[htbp]
\centering
\includegraphics[width=9.5cm,height=7cm]{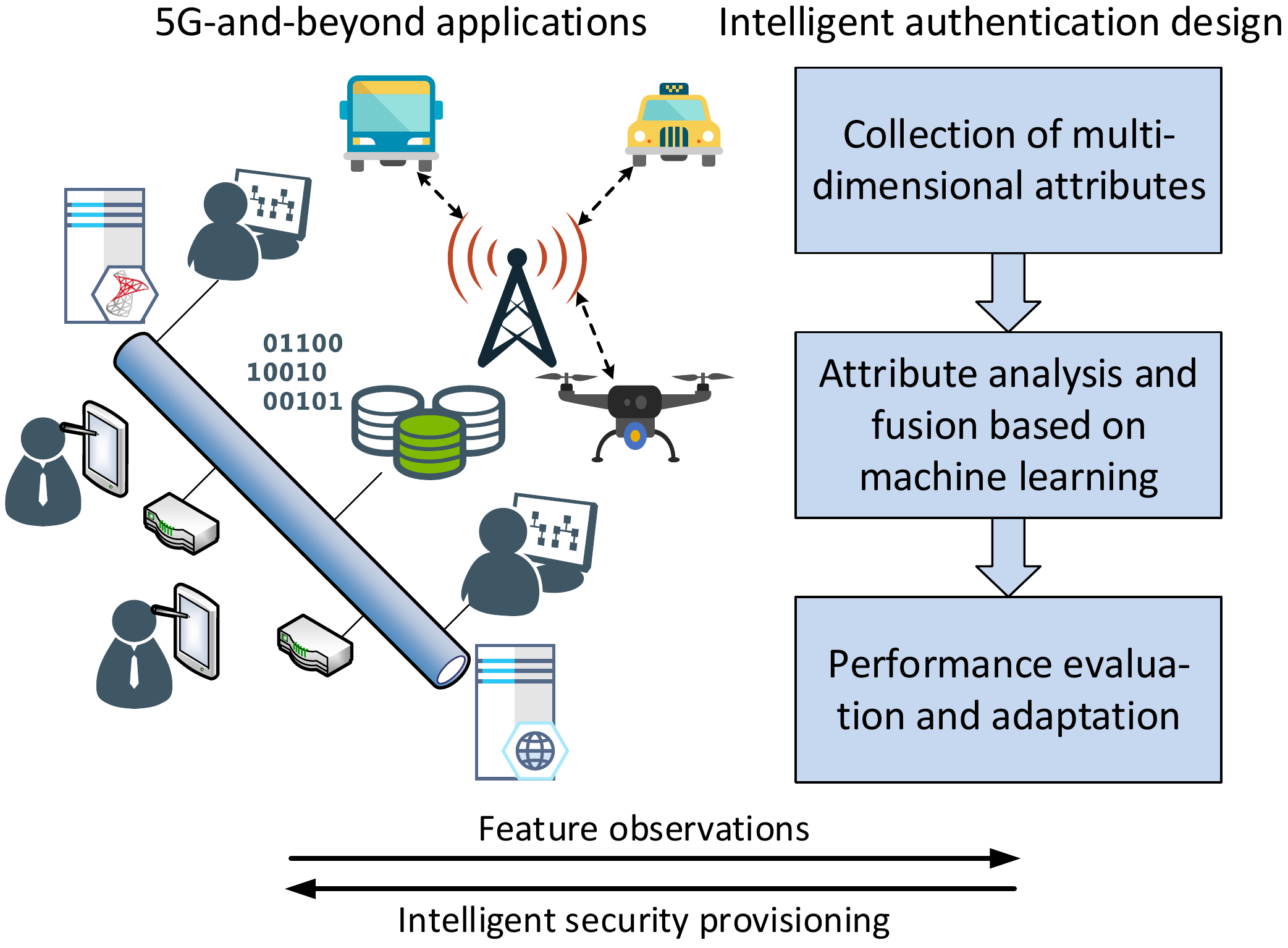}
\caption{Framework diagram of intelligent authentication design.}
\end{figure}\\
\emph{Phase I:} The time-varying multi-dimensional attributes are collected for authentication, which may be estimated imperfectly having noises and measurement errors. 
Examples include the physical layer attributes, network selection in heterogeneous wireless networks, and mobility patterns. In a specific 5G-and-beyond wireless communication scenario, those attributes providing more information for authentication may be selected first. In details, the independent attributes having a broader distribution range and a higher estimation accuracy could offer more information for distinguishing different transmitters.
By utilizing multi-dimensional attributes as well as sharing the information among different layers and networks,  the \emph{reliability} of authentication will be improved. Explicitly, the design of intelligent authentication only relies on the
estimation data of attributes without requiring a precise structure of the time-varying attributes, e.g., the channel model, resulting in a \emph{model-free} device validation.\\
\emph{Phase II:} The  multi-dimensional attributes can be fused for authentication based on machine learning techniques. An example is given in our previous work \cite{01}, which is a kernel machine learning-based physical layer authentication scheme. More machine learning paradigms for intelligent authentication will be summarized shortly. Considering the time-varying network conditions, such as the resource limitations and uncertainties, the attributes may be opportunistically selected for dealing  with both communication overhead and security management concurrently. Furthermore, developing an appropriate machine learning algorithm and reducing the dimension  of authentication system also benefit   the communication  performance, thus \emph{cost-effective} authentication will be achieved. \\
\emph{Phase III:} The authentication for Alice and the Spoofer can be conducted based on the new collection of multi-dimensional attributes. To achieve this, the regression or classification model should be built based on the training data collected from Alice/Spoofer.
  Then the authentication performance can be evaluated, and the authentication process can be adapted to the complex time-varying environment by exploring machine learning to track the variations of multi-dimensional attributes.
Hence, the \emph{continuous} and \emph{situation-aware} process is proposed for intelligent security provisioning in 5G-and-beyond applications.

\section*{MACHINE LEARNING PARADIGMS FOR INTELLIGENT AUTHENTICATION}

The family of machine learning algorithms can be categorized based on  their functionality and structure \cite{31}, yielding regression algorithms, decision tree
algorithms, Bayesian algorithms, clustering algorithms, and artificial neural networks, just to name a few. In this article,
we clarify the machine learning techniques from two perspectives: parametric/non-parametric learning and  supervised/unsupervised/reinforcement learning, as illustrated in Fig. 4.
The adjectives ``parametric/non-parametric" indicate whether there are specific forms of training functions, while ``supervised/unsupervised" indicate whether there are labeled samples in the database. The focus of  reinforcement learning  is finding a balance between exploration of uncharted territory and the exploitation of current knowledge.
We will introduce the basic concepts and typical examples of these machine learning techniques, more importantly, discuss their applications and possible requirements in designing intelligent authentication approaches  for different wireless communication scenarios.
\begin{figure}[htbp]
\centering
\includegraphics[width=15.5cm,height=5cm]{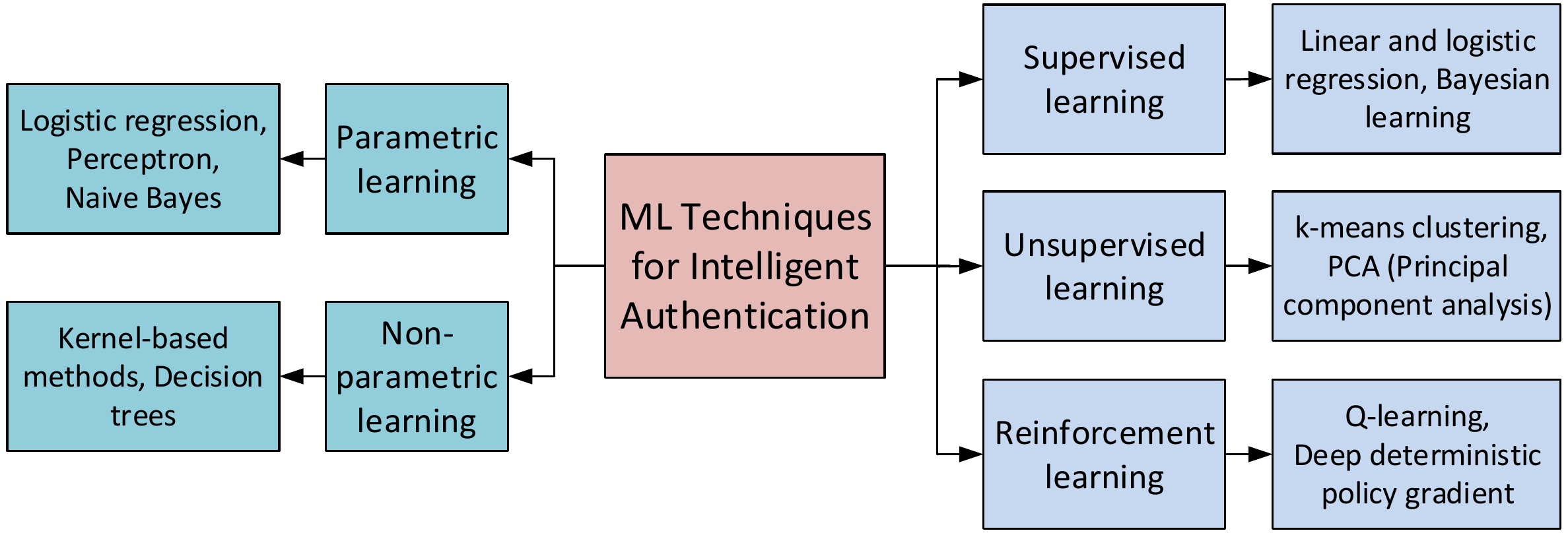}
\caption{Categories of machine learning (ML) techniques for intelligent authentication.}
\end{figure}\\
\textbf{Parametric learning methods:} The family of parametric learning methods has
become mature in the literature,  as exemplified by the logistic regression, linear discriminant analysis, perceptron, and naive Bayes, which require the specific form of training functions \cite{32}. When the training functions related to the training samples (i.e., the collection of multi-dimensional information) are selected appropriately, the parametric learning methods could be more accurate,
simpler, and require less training samples than the non-parametric learning methods.
For instance, the authors of \cite{87} proposed a logistic regression-based authentication scheme by utilizing the RSSI of transmitters for authentication enhancement  assuming  that all the radio nodes are static.

In the intelligent authentication schemes, the parametric learning methods could model the attributes independently based on the specific form of training functions, so that the uncertainties caused by the complex time-varying environment may be circumvented, as well as the communication overhead and complexity of training may be adjusted adaptively by opportunistically leveraging  attributes. However, this kind of learning methods may be limited in 5G-and-beyond wireless communication scenarios wherein the statistic properties and a priori knowledge of attributes are not at hand.\\
\textbf{Non-parametric learning methods:}  In contrast to parametric learning methods, the non-parametric learning methods \cite{01,31,32} are not specified \emph{a priori}, but are determined from the available data. Examples include  kernel estimator, k-nearest neighbors, and decision trees, just to name a few.  A kernel machine-based scheme for intelligent physical layer authentication process is proposed in \cite{01}. The dimensionality of a multiple attribute-based authentication system
is reduced by the kernel function and the resultant authentication process can be modeled as a linear system, thus decreasing  the computation complexity of
authentication process, even though a large number of attributes are utilized.
More importantly, the proposed kernel learning algorithm tracks the time-varying attributes to enhance security  based on continuous device validation.

The non-parametric learning methods  learn dynamically from time-varying environments without requiring any assumptions concerning the training models. Beneficially, this provides higher flexibility for the intelligent authentication, especially in those real-time authentication scenarios, where the computational resources and time available for obtaining the statistical
properties of attributes and training functions are limited. However, compared with the parametric learning methods, they require more training data (i.e., collection of multi-dimensional information and/or their corresponding labels) and may result in overfitting.\\
\textbf{Supervised learning algorithms:} The main difference between  supervised and unsupervised learning algorithms is that supervised learning requires the prior knowledge of outputs for the corresponding inputs, while unsupervised learning algorithms do not need labeled outputs. Some supervised machine learning algorithms are studied in \cite{32} for defending against the false data injection attacks, such as the perceptron, k-nearest neighbors, and support vector machines (SVM).
In intelligent authentication schemes, the choice between supervised or unsupervised machine learning algorithms typically depends on the authentication problem and volume of training data at hand. A supervised learning algorithm is suitable when training data and corresponding outputs of a legitimate communication session are straightforward to obtain. More explicitly, the labeled outputs should be required near-instantaneously, so that the adaptation of the authentication process may be achieved in real time. When labeled outputs are not at hand, fast labeling techniques may be helpful for the supervised learning-based intelligent authentication.\\
\textbf{Unsupervised learning algorithms:} In exploring this kind of algorithms, the learner (Bob) exclusively receives unlabeled training data and makes predictions for all unseen points, as exemplified by the K-means clustering~\cite{31}.
In scenarios wherein the samples of Alice (estimates of attributes) are much more than that of Spoofer, we may introduce unsupervised learning algorithms for intelligent authentication.  It is reasonable to assume that the observations of Alice are much more than that of Spoofer, since the spoofing attacks fully rely on the  Alice's behaviors for better attacking performance, including the transmission protocols and attributes. Through applying unsupervised machine learning techniques for identifying Alice, the time latency and energy consumption for labeling the outputs are significantly decreased  in the intelligent authentication process.\\
\textbf{Reinforcement learning:}  This kind of techniques does not require accurate inputs and outputs as well as precise parameter updates. A Q-learning-based authentication scheme is proposed in \cite{85} depending on the RSSI of signals to achieve the optimal test threshold and to improve the authentication accuracy. However, it  is also a static authentication scheme, and may be unsuitable when the available resources and time are limited for obtaining complete information, i.e., the
environment and agent states as well as the immediate reward for each action of devices.

To improve the  authentication performance, we may utilize multi-dimensional attributes in higher layers (e.g., application layer) and explore more precise information of Alice, but her privacy may not be guaranteed. In other words, Bob may collect Alice's information, such as user behaviors and location-related features, for analyzing her habits, locations, and other sensitive information during the authentication process, thus leading to privacy leakage \cite{86}.
Hence, developing a privacy preserving authentication scheme based on the masking methods~\cite{21} is helpful in protecting Alice's private information during the authentication process by adding obfuscation patterns on attribute measurements. However, there is a trade-off between   privacy and  data utility \cite{21}, thus resulting in a trade-off between privacy preservation and authentication performance. To be specific, by adding obfuscation patterns on attribute measurements, the information/contexts may be released to provably preserve Alice's privacy, while the authentication performance relying on the reduced accuracy of attribute measurements will be decreased.

In order to explore a better trade-off between privacy preservation and
authentication performance, we may explore the masking-based privacy protection technique with spatial and temporal aggregation of attributes' measurements  for intelligent authentication.
Such technique generates and aggregates obfuscation patterns for each attribute on Alice side at a specific time instant $t$ as $\overline{\bm{H}}[t]=\bm{H}[t]+\bm{\gamma}[t]$, where $\bm{H}=(H_{1},H_{2},...,H_{N})^{\rm{T}}$ and $\bm{\gamma}=(\gamma_{1},\gamma_{2},...,\gamma_{N})^{\rm{T}}$ represent the estimates of $N$ attributes used and generated obfuscation patterns, respectively. More importantly, the obfuscation patterns are generated to ensure that
the fusion of $\overline{\bm{H}}[t]$ for authentication by machine learning should be near to the fusion of original attribute estimates $\bm{H}[t]$.
In achieving this, information sharing between different layers of network architecture  is required. When an attribute  is not privacy-sensitive for Alice, the obfuscation pattern superposed on this attribute should be set to zero, so that the system capacity is guaranteed.
As a conclusion, through the masking-based privacy protection technique with the spatial aggregation and temporal aggregation,  each sensitive attribute of Alice has all properties of random number series, thus reveals no information to Bob.

\section*{PERFORMANCE ANALYSIS}

A detailed comparison of conventional physical layer authentication and machine learning-aided intelligent authentication techniques is given in Table I. To be more specific, differently from the conventional techniques \cite{06,02,72}, machine learning-aided intelligent authentication techniques make full use of both varying attributes and adaptive authentication processes, and implicitly  learn from data without the requirement of an exact attribute model. More importantly, the machine learning-aided intelligent authentication techniques achieve cost-effective, more reliable, model-free, continuous, and situation-aware device validation  in the complex time-varying environment, as exemplified by \cite{01}. The cost is that they  usually require the appropriate selection of machine
learning algorithms and elaborate design of authentication processes, especially for delay-sensitive communications and resource-constraint devices.

\begin{figure}[htbp]
\centering
\includegraphics[width=16cm,height=4.6cm]{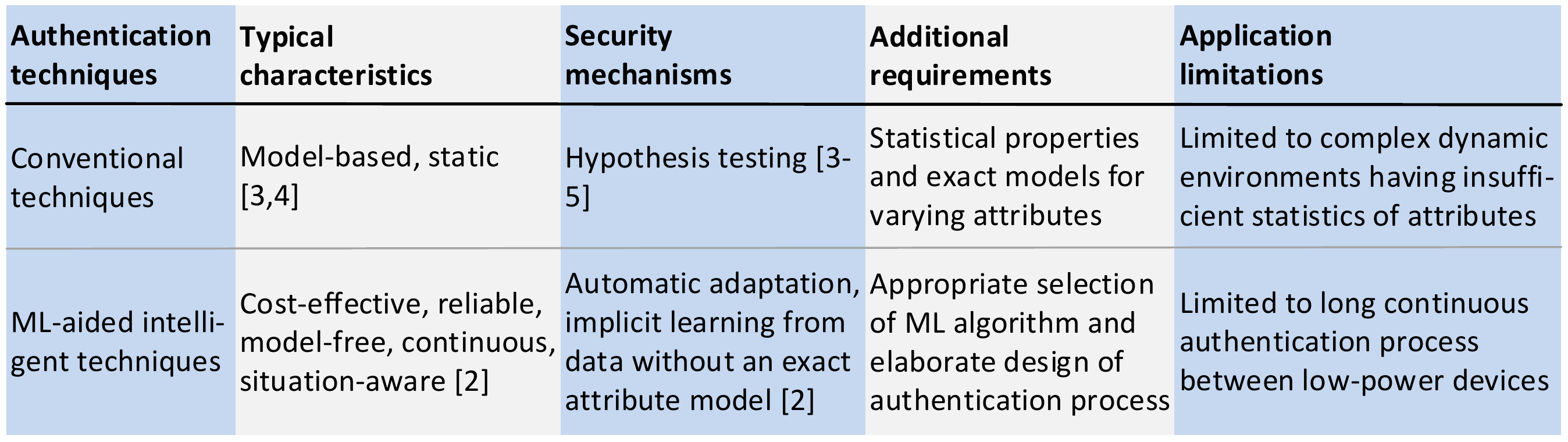}
\caption*{TABLE I: Comparison between the conventional physical layer authentication and machine learning (ML)-aided intelligent authentication techniques in complex time-varying 5G-and-beyond wireless networks. }
\end{figure}

\begin{figure}
\centering
\subfigure[Authentication performance comparison between the intelligent authentication scheme and the static authentication scheme. ]{
\includegraphics[width=9.5cm,height=8cm]{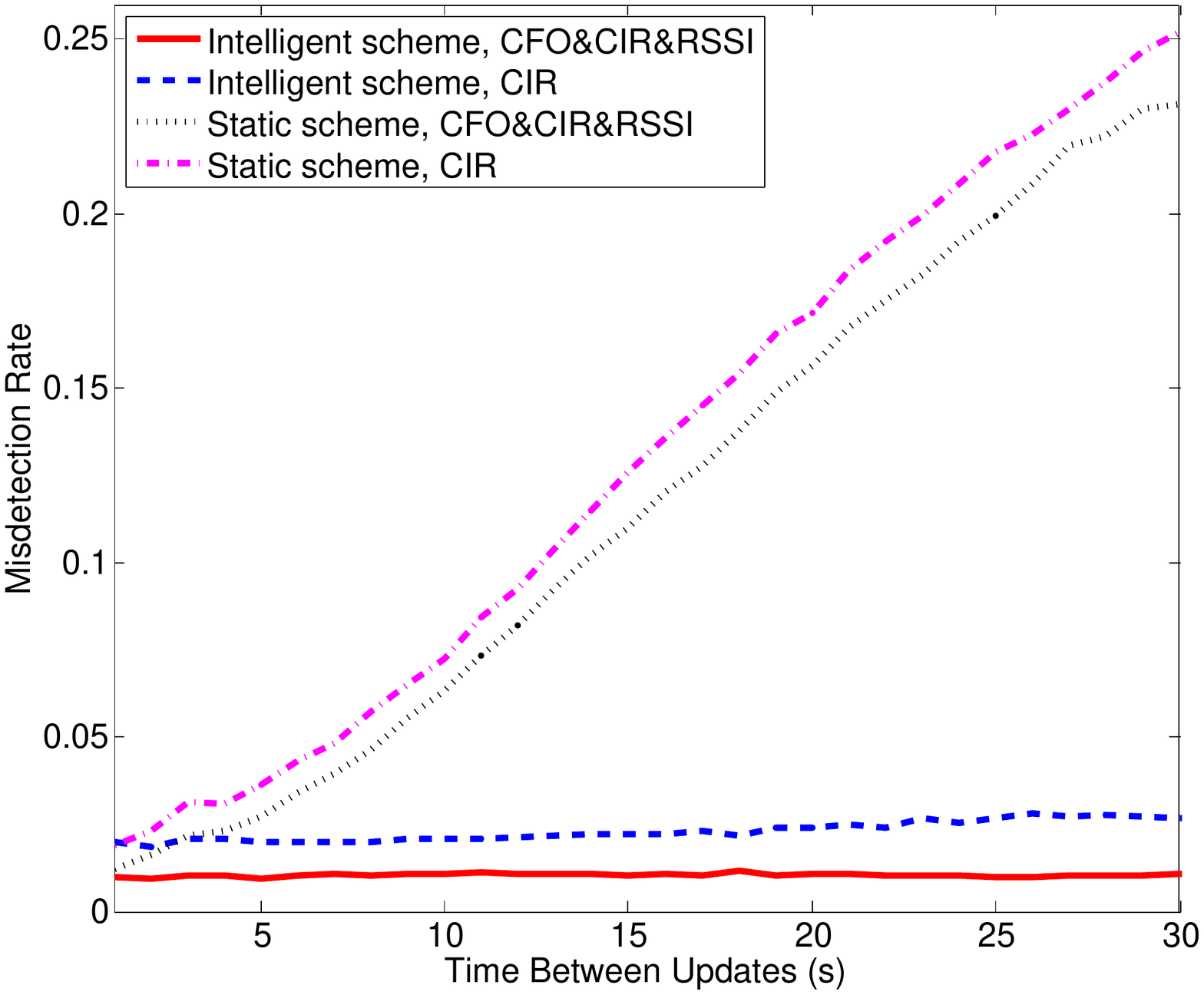}}
\subfigure[Computation cost comparison between the intelligent authentication scheme and the schemes of \cite{85}. ]{
\includegraphics[width=9.5cm,height=8cm]{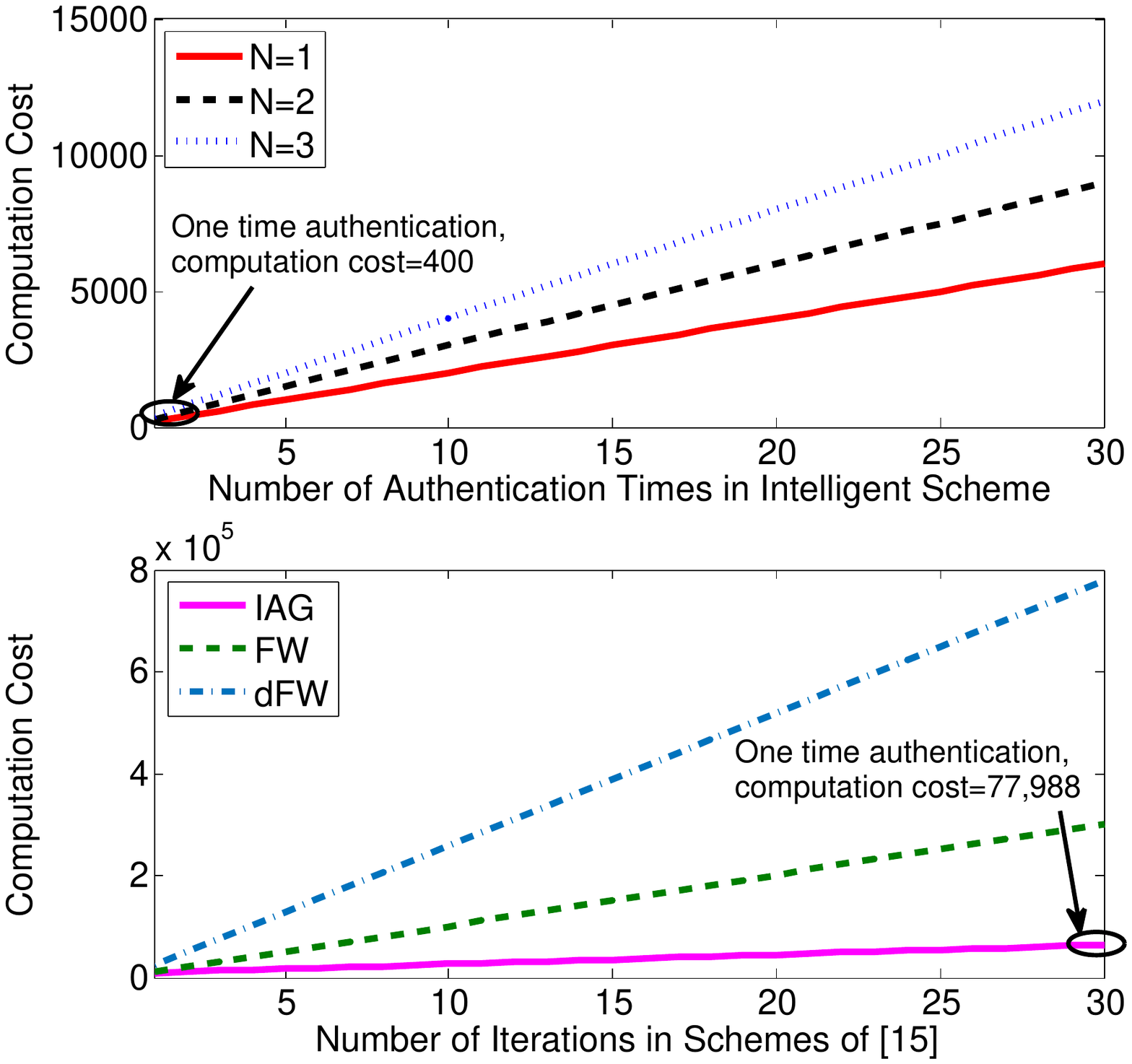}}
\caption{Performance of the developed intelligent authentication approach.}
\end{figure}

To demonstrate  the performance of the developed  intelligent authentication approach, Fig. 5 (a)  compares the simulation results of the kernel machine learning-based intelligent authentication scheme and the static authentication scheme relying on both multiple attributes (i.e., CFO, CIR, and RSSI) and single attribute (i.e., CIR).  The attribute estimates are generated based on the existing works of  \cite{01,02,06,72}, and the attribute estimates of the Spoofer are not required for training  this intelligent authentication scheme, since it tracks Alice's attributes and authenticates the estimates that are different when coming  from Alice and the Spoofer. Estimates of Spoofer's attributes for deriving misdetection rate are randomly generated within the estimation ranges. We can observe from Fig. 5 (a) that both schemes achieve a lower misdetection rate by exploring multiple  attributes. The reason for this trend is that  the authentication accuracy can be increased at legitimate devices in complex time-varying environments through exploring multiple attributes, since it is more difficult for the Spoofer to predict and imitate multiple attributes of Alice.
It is also observed from Fig. 5 (a) that upon increasing the time between updates, the misdetection rate of the intelligent authentication scheme remains stable and robust, while that of the static authentication scheme increases dramatically. This demonstrates that without the situation-aware adaptation, the performance of the static authentication scheme will be dramatically decreased in complex time-varying environments, thus limiting its application in 5G-and-beyond wireless networks. In order to achieve continuous authentication, the static authentication scheme has to repeat the authentication process by collecting attribute estimates,  obtaining the statistical properties of the attributes, deriving new test threshold, and then reauthenticating the devices. Through repeating this process, the static authentication scheme requires more computation resources and longer time latency.

 Fig. 5 (b) characterizes the computation cost of both  the developed kernel machine learning-based intelligent scheme and the schemes in \cite{85}, namely for incremental aggregated gradient (IAG), Frank-Wolfe (FW), and distributed Frank-Wolfe (dFW) algorithms. Note that the abscissa of the first figure indicates the number of  authentication times while that of second figure represents the iteration number in  the schemes of \cite{85}.
Explicitly, the schemes of \cite{85} are static and one-time authentication solutions requiring  the help of landmarks for collecting RSSI of signals, while the developed intelligent  scheme utilizing $N$ attributes, i.e., $N=1$, $N=2$, and $N=3$, is continuous and situation-aware. The computation cost of one time authentication in the developed intelligent scheme is only 400, while the schemes of \cite{85} require many iterations to reach optimal solutions, thus leading to much higher computation cost, namely for 77,988.
It can also be observed from Fig. 5 (b) that although the larger number of authentication times  increases the computation cost in the developed intelligent scheme, it provides continuous adaptive authentication and enhanced security for legitimate devices.

\section*{CONCLUSIONS}

This article firstly introduced the challenges for the conventional  authentication techniques and the advantages of intelligent authentication. The intelligent authentication design was developed to enhance security performance in 5G-and-beyond wireless networks. Then the machine learning paradigms for intelligent authentication were classified into parametric and non-parametric learning methods, as well as supervised, unsupervised, and reinforcement learning techniques. As a conclusion, machine learning techniques provide a new insight into authentication under unknown network conditions and unpredictable dynamics, and bring intelligence to the security management to achieve cost-effective, more reliable, model-free, continuous, and situation-aware authentication.

\section*{FUTURE WORK}
In 5G-and-beyond networks,  there are also many other research  areas in which machine learning can play a remarkable role and improve the quality of services. We summarize a range of future research ideas on machine learning for  intelligent  security and  services as follows.

The machine learning techniques may be utilized for other security applications, such as anomaly/fault/intrusion detection, access control, and authorization. This is mainly due to their ability to provide continuous protection for legitimate communications in 5G-and-beyond networks.
Furthermore, with the ongoing convergence between wireless devices and human beings, machine learning provides a new insight for studying human-device interaction, and the interplay between devices and information security, as well as the database security and data mining, operation systems security, Internet and cyber-security, incident handling, hacking, biometric techniques, smart cards, infrastructure protection, and risk management. Through identifying and learning the dynamic adversarial systems, the automatic security management may be achieved by machine learning techniques.

With the fast development of distributed communication  systems (e.g., blockchain), machine learning may facilitate distributed security management.
To be specific, it may be explored  for inferring the mobile users' decision making and device's dynamic states under unknown network conditions for better security performance, for example, adversarial behaviors study for predicting the possible attacks of adversaries in peer-to-peer networks.
The family of machine learning algorithms may be also applied for better decision making, such as resource allocation, distributed computing,
analytical thinking, customer orientation, strategy, and planning. These are expected to be extremely important for 5G-and-beyond wireless networks to achieve intelligent and autonomous services for human beings.

\ifCLASSOPTIONcaptionsoff
  \newpage
\fi

\appendices

~\\

\section*{Biographies}~\\
\textbf{HE FANG} is a Ph.D. candidate at the Department of Electrical and Computer Engineering, Western University, Canada. She received her B.Sc. and Ph.D. degrees in Applied Mathematics from Fujian Normal University, China, in 2012 and 2018, respectively. Her research interests include intelligent security provisioning, machine learning, as well as distributed optimization and collaboration techniques.
One focus of her current research is on the development of new machine-learning enabled authentication schemes through utilization of time-varying wireless environment for security enhancement. She is also working on distributed security management in IoT and blockchain systems under practical network constraints.~\\
\textbf{XIANBIN WANG} [S'98-M'99-SM'06-F'17] is a Professor and Tier 1 Canada Research Chair at Western University, Canada. His current research interests include 5G technologies, Internet-of-Things, communications security, machine learning and intelligent communications. Dr. Wang has over 380 peer-reviewed journal and conference papers, in addition to 29 granted and pending patents and several standard contributions.
Dr. Wang is a Fellow of Canadian Academy of Engineering, a Fellow of IEEE and an IEEE Distinguished Lecturer. He has served as Editor/Associate Editor/Guest Editor for more than 10 journals. Dr. Wang was involved in over 50 conferences with different roles such as symposium chair, keynote speaker, tutorial instructor, track chair, session chair and TPC co-chair.~\\
\textbf{STEFANO TOMASIN} (tomasin@dei.unipd.it) is with the Department of Information Engineering of the University of Padova, Italy. His current research interests include physical layer security and signal processing for wireless communications, with application to 5th generation cellular systems. In 2011-2017 he has been Editor of the IEEE Transactions of Vehicular Technologies and since 2016 he is Editor of IEEE Transactions on Signal Processing. Since 2011 he is also Editor of EURASIP Journal of Wireless Communications and Networking.

\end{document}